\documentclass[twocolumn,aps,prb,showpacs,preprintnumbers,amsmath,amssymb, superscriptaddress]{revtex4}

\usepackage{graphicx}
\usepackage{dcolumn}
\usepackage{bm}

\begin{document}


\title{Percolation through Voids around Overlapping Spheres, a Dynamically based Finite Size Scaling Analysis} 


\author{D. J. Priour, Jr}
\affiliation{Department of Science, Kansas City Kansas Community College, Kansas City, KS 66112, USA}


\date{\today}

\begin{abstract}
The percolation threshold for flow or conduction through voids surrounding 
randomly placed spheres is rigorously calculated.  With large scale 
Monte Carlo simulations,
we give a rigorous continuum treatment to the geometry of the impenetrable spheres and 
the spaces between them.  
To properly exploit finite size scaling, we examine multiple systems of differing sizes,
with suitable averaging over disorder, 
and extrapolate to the thermodynamic limit.
An order parameter based on the statistical sampling of stochastically driven 
dynamical excursions and amenable to finite size 
scaling analysis is defined, calculated for various system sizes,
 and used to determine the critical 
volume fraction $\phi_{c} = 0.0317 \pm 0.0004$ and 
the correlation length exponent $\nu = 0.92 \pm 0.05$.
\end{abstract}

\maketitle


\section{Introduction}

Notwithstanding simple underlying Physics, 
percolation transition are \textit{bona fide} second order 
phase transitions, with all of usual the hallmarks of singular 
behavior at a critical point~\cite{stauffer}.  Percolation transitions, 
phase transitions mediated by a disordering influence such as the    
random removal of sites or bonds on a regular lattice 
geometry, are amenable to numerical study in the context 
of Monte Carlo calculations used to treat disorder.  However, 
while the critical behavior in discrete systems may be 
characterized to a high level of precision in this manner, there are salient 
examples of percolation phenomena which cannot be reduced to discrete 
lattices.

In realistic porous media, the flow of liquid on a macroscopic
basis typically entails the passage of liquid around irregularly shaped
granules which comprise the material and represent barriers to 
fluid flow.
For the movement of fluid or charge in porous materials, there 
is a dichotomy for volume concentrations of barrier 
particles above and below a critical concentration $\rho_{c}$.
If inclusions are sufficiently sparse, voids join together 
in a macroscopic connected network spanning the entire 
system, allowing for the transmission of fluid at the bulk 
level.  On the other hand, for large enough concentrations of 
randomly placed barrier particles, contiguous navigable voids are  
limited in size to a finite length scale $\xi$, barring 
the transmission of fluid at a system wide level; the 
critical density $\rho_{c}$ of inclusions which separates 
the two scenarios is thus of    
practical significance.  

For the sake of convenience, we rescale coordinates in such a way 
that the spherical inclusions are each of unit radius.  Although in 
principle one may record the critical density $\rho_{c}$ 
where the percolation transition occurs, a more frequent practice 
in discussing percolation through voids 
is to report instead the excluded volume~\cite{zallen}, given by
 $\phi_{c} = e^{-\frac{4 \pi}{3} \rho_{c}}$.

Determining if voids 
may be traversed at the macroscopic level is a continuum
percolation problem for which a topologically equivalent 
discrete counterpart is in general not readily accessible, a 
condition very distinct from that of a complementary 
percolation problem (with the permeability pattern reversed) 
where the flux occurs through networks 
formed of overlapping included particles with interstitial 
regions impervious to fluid flow.  
The geometry of the randomly placed  
barriers, known \textit{a priori}, facilitates the calculation
of characteristics related to percolation when the flow is   
through included particles instead of the interstitial 
spaces.  On the other hand, the fact that the shapes of voids 
between barrier particles are only evident \textit{a posteori}
hampers a rigorous identification of networks formed 
by connected voids and enhances the challenge of finding the   
percolation threshold.
Nevertheless, with the aid of a finite size scaling analysis, we determine 
not only the critical volume fraction $\phi_{c}$, but also the 
critical exponent $\nu$ associated with the correlation length $\xi$.

Previous theoretical studies include calculations which use an approximate discrete 
scheme and extrapolate to the continuum case~\cite{torquato,maier}.  In a related effort, the 
discretization scheme has been applied to randomly placed ellipsoids~\cite{yi} 
and oblate spheroids~\cite{yi2}. 
In particular cases, such as systems comprised of randomly placed spheres,
discrete networks equivalent to void spaces~\cite{kerstein} may be constructed.
Voronoi tessellations, and the appropriate generalizations have been 
applied~\cite{kerstein0,marck,rintoul} to systems comprised of 
randomly located spheres.
In a recent set of 
calculations, there is no 
discretization scheme and the simulation entails following tracer particles 
in an effectively infinite sized system~\cite{hofling,spanner}; the
percolation threshold is identified by
determining the density of spheres for which the tracers cease to move
diffusively.
In this work, we use a dynamical approach where systems of finite 
size are examined.  In this way, we are able to apply standard machinery of 
finite size scaling analysis while providing rigorous treatment 
to the geometric intricacies of voids between barriers which serve
as conduits for fluid or charge in the percolative regime.

\section{Calculation of the Excursion Order Parameter}

An order parameter, an intensive 
thermodynamic variable finite when the phase is intact, and decreasing 
continuously to 
zero at a phase boundary (e.g. magnetization per unit volume in the 
case of ferromagnetism disrupted by thermal fluctuations above the 
Curie Temperature and nonzero below it),
may be pressed into service as a tool to locate second order phase 
transitions and thereby determine the phase diagram.
In percolation phenomena where connected navigable clusters 
are readily identified, the ``strength'' of the spanning 
cluster, or the number of sites in the lattice it encompasses 
relative to the total number of sites in the system, is 
conveniently sampled in Monte Carlo simulations using 
techniques such as the Hoshen-Kopelman algorithm for 
the identification and characterization of connected clusters~\cite{hoshen}. 

In the scenario of interest, where percolation occurs through 
irregularly shaped interstices among randomly placed barrier particles of a prescribed 
shape, the connectivity of void volumes to adjacent navigable 
regions may be difficult to establish in more general situations
(e.g. non-spherical inclusions such as tetrahedra, cubes, or octahedra).
To address this challenge, we use dynamical simulations to probe and 
determine the extent of voids.  Although stochastically driven exploration is 
validated in the case of randomly placed spheres, the approach
may be used in scenarios where variants of Voronoi tessellation 
are more difficult to apply.

In the spirit of a \textit{Gedanken} experiment,
we envision applying a reflective coating to 
exposed surfaces of overlapping spheres such that 
light striking the included particles is 
reflected with no penetration of the incident rays.  
Whereas the transmission of light through the 
matrix of inclusions would signal that the system would 
admit the flow of fluid or electrical charge, a configuration 
which blocks the transmission of incident light also
lacks a percolating path.

Dynamical trajectories are initiated with an unbiased choice 
for the point of origin in the interior of a void and 
exterior to all barrier particles.  To select a void 
point, randomly chosen candidate locations are accepted 
only if the point is exterior to all neighboring spheres.
Some economy is gained in 
partitioning the 3D system into a cubic grid, where the unit cell     
size is slightly greater than the radius $r$ of the spheres.   
The subdivisions reduce the task of checking for penetration 
into nearby spheres to simply checking the cubic unit cell occupied by the        
prospective (randomly chosen) starting point as well as each of the 
adjacent cubes.  To set the virtual ray in motion,
a ``velocity'' vector $\hat{v}$ is randomly chosen in a spherically symmetric 
way to eliminate any directional bias, with $\hat{v}$ of unit length since ``speed'' has no physical or computational 
significance in the 
context of the order parameter calculation.

The ray trajectory is $\vec{x}_{l} = \vec{x}_{0} + \hat{v} t$ with 
the starting point $\vec{x}_{0}$ and the orientation of the velocity vector $\hat{v}$ stochastically determined.
The path of the ray must be traced until contact is made with the 
nearest sphere; with a barrier particle centered at $\vec{x}_{c}$, the condition 
for intersection with a sphere of radius $r$ is $|\vec{x}_{l} - \vec{x}_{c} | = r$.  Standard 
geometric arguments lead to $d_{\mathrm{coll}} = -x_{d} \cos \theta \pm \sqrt{1 - x_{d}^{2} \sin^{2} \theta}$
for the distance to a point of intersection, 
where $x_{d} \equiv |\vec{x}_{0} - \vec{x}_{c} |$ and $\cos \theta \equiv \frac{1}{x_{d}} \vec{v} \cdot \vec{x}_{d}$.
Choosing the negative sign corresponds to the closer point of entry and the positive sign to the more 
distant point of departure as the ray propagates through and exits from the sphere.   
Finding the closest included particle intersected by the ray is 
tantamount to minimizing $d_{\mathrm{coll}}$,  
the shortest distance to a sphere along the trajectory. 

To reduce computational overhead, the system is subdivided into a cubic grid in the 
same fashion as used to determine if a randomly chosen point is located in a void; rays are 
propagated from one small unit cell to the next until a collision with a barrier particle is 
recorded.  At the site of a collision $\vec{x}_{\mathrm{coll}}$, to avoid penetration of 
the impermeable particles comprising the matrix, the direction of 
motion must change in such a way that the incident ray is projected away from the interior of the
barrier particle.  Specular reflection is implemented by reversing the component of the velocity 
in the direction of $\vec{x}_{\mathrm{coll}} - \vec{x}_{c}$, the radius vector extending 
from the sphere center to the site of the collision.  

     Although mirror reflection encapsulates the physics of geometric optics appropriate 
to a ray of light moving in a matrix of silvered spheres, propagation by simple 
specular reflection may be improved on, particularly in the case of neck-like voids where many 
collisions may be required to traverse the narrow corridor if the ray is directed normal to the 
primary axis of the void.     To improve the 
efficiency of navigation by roughly an order of magnitude, we incorporate moves where, instead of mirror 
reflection, the ray trajectory shifts by a right angle. In 3D, 
even with a perpendicular shift imposed, there remains a degree of 
freedom, which is selected at random 
to provide stochastic input at each collision.  The balance of the moves is 
such that stochastically infused right angle moves outnumber specular moves two to one. 

To interrogate the system and determine whether a void is part 
of a spanning cluster, we 
calculate the excursion parameter $\tilde{\delta} = \frac{1}{3L} (\delta_{\mathrm{rms}}^{x} 
+ \delta_{\mathrm{rms}}^{y} + \delta_{\mathrm{rms}}^{z})$, structured to 
allow for the incorporation of statistical information 
gleaned from traversals along each of the three Cartesian 
axes.  In calculating the rms deviations, special consideration must be given
the periodicity of the system.  When the density of randomly placed
spheres is high enough to block percolation, where $\phi < \phi_{c}$, the scale $\xi$ of
contiguous void networks is in most cases small relative to the system size, and 
accordingly, $\tilde{\delta} \ll 1$.

On the other hand, if inclusions are sparse enough that $\phi > \phi_{c}$,
system wide traversal becomes feasible.
Moreover, with periodic boundary conditions imposed the RMS deviations along one or 
more axes could in principle grow without bound as the ray propagates from one replica
of the system to the next.  In such situations, the result would be an  
unrealistically large order parameter, a circumstance which may be remedied 
by truncating $\delta_{\mathrm{rms}}^{x}$ and counterparts 
along other axes at the system size $L$.  Thus, in the extreme case
of a system devoid of inclusions, $\tilde{\delta}$ would attain its 
maximum value of 1. 
The excursion quantity $\tilde{\delta}$, which saturates at unity 
in the absence of obstacles and falls to zero at the 
percolation threshold for bulk systems, serves as an    
order parameter giving a measure of the size of a randomly
selected void. 
In the thermodynamic limit, the configuration averaged $\tilde{\delta}$ also gives the 
probability that a randomly selected void is a system spanning cavity.

In a diffusive excursion, though the disorder and time 
averaged vector displacement vanishes, the root mean square 
distance traveled scales as $N_{\mathrm{coll}}^{1/2}$, where $N_{\mathrm{coll}}$ 
is the number of ray interactions (and concomitant 
trajectory deflections) with barrier particles.
Unless one is precisely at the percolation threshold with 
RMS deviations from the trajectory starting point scaling as
$N_{\mathrm{coll}}^{\epsilon}$, where $\epsilon < 1/2$ in the non-diffusive 
critical regime~\cite{hofling,spanner}, configuration averaged ray 
excursions will scale diffusively with the number of sphere interactions,  
if the system is sufficiently large.

If the void is small in size, a ray which propagates in the cavity will probe its
extent relatively quickly, yielding a finite $\tilde{\delta}$ value.  On the other hand,
with a system spanning void scaling as the system size $L$, the ray traversal
time varies as $L^{2}$ on average.  Hence, to avoid a spuriously low 
result for the percolation threshold $\phi_{c}$, trajectories are propagated 
with collisions accumulating until $N_{\mathrm{coll}} \gg \eta L^{2}$ where $\eta$
is an integer prefactor.  By successively doubling $\eta$, a saturation point 
is eventually found (i.e. for $\eta = 80,000$ and $\eta = 160,000$) 
where statistically significant changes in 
critical quantities cease to occur.

The calculation of the $\tilde{\delta}$ parameter entails averaging
over many realizations of disorder (i.e. at least $10^{4}$) to 
suppress statistical fluctuations and
permit the observation of finite size scaling trends, which are exploited 
to calculate critical quantities such as $\phi_{c}$ and the correlation 
length exponent $\nu$.  To generate disorder configurations, we sample 
the grand canonical ensemble as described in the Appendix to determine the  
number of spherical barriers encompassed in the simulation volume.
The Mersenne Twister algorithm, which generates high quality 
random numbers with a modest computational cost, is used for stochastic 
input.

\begin{figure}
\includegraphics[width=.49\textwidth]{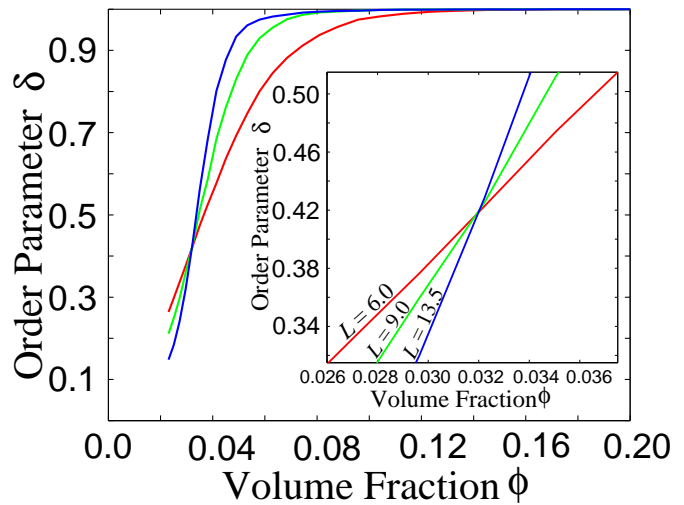}
\caption{\label{fig:Fig1} (Color online) The excursion order 
parameter $\tilde{\delta}$ is plotted with respect to excluded
volume $\phi$ for various system sizes; the inset is a magnified view 
of the intersection of the $\tilde{\delta}$ curves.}
\end{figure}

\section{Finite Size Scaling Analysis}

Percolation transitions, whether in the context of continuum or in 
discrete cases, are second order transitions with singular behavior in 
thermodynamic variables at the phase transition.  Single parameter finite 
size scaling analysis posits a form $L^{\alpha} f([\phi-\phi_{c}] L^{1/\nu})$ for thermodynamic variables, a 
singular form anticipated for the excursion order parameter $\tilde{\delta}$. 
Since $\tilde{\delta}$ is normalized and thus dimensionless, $\alpha = 0$, and the
singular dependence reduces to $\tilde{\delta} \sim f([\phi - \phi_{c}] L^{1/\nu})$.  An important  
implication is the crossing of $\tilde{\delta}$ curves at $\phi_{c}$ for different system sizes, 
provided the system size is large enough. Similar intersections are seen in the case of the Binder 
cumulant~\cite{binder} at the Curie temperature  $T_{c}$ in the context of ferromagnetic 
transitions, and the crossing phenomena in both the former and the latter represent
a robust avenue for locating 
second order phase transitions.

\begin{figure}
\includegraphics[width=.35\textwidth]{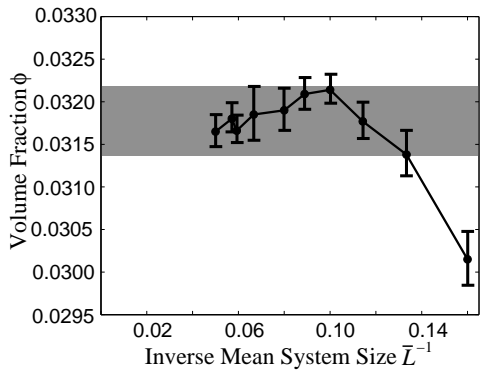}
\caption{\label{fig:Fig2} (Color online) Pseudocritical excluded volumes
(with calculated error bars)
plotted with respect to the reciprocal of the mean system size $\bar{L}$ 
for each pair}
\end{figure}

Fig.~\ref{fig:Fig1} shows $\tilde{\delta}$ curves for different system sizes (i.e. $L = 6.0$, $L = 9.0$, and $L = 13.5$)
with a well defined crossing at a common point.  To determine $\phi_{c}$ in a systematic fashion, we calculate
the order parameter for pairs of system sizes 
(e.g. $\left \{ 6,9 \right \}$ and $\left \{8, 12 \right \}$) with the larger       
member of each pair 1.5 times the size of the smaller.  
Although in the thermodynamic limit crossings would occur exactly at $\phi_{c}$,
the locations of intersections for finite size systems [e.g. $\phi_{c}(\bar{L})$ differ 
slightly from $\phi_{c}$ due to finite size effects, and must be considered pseudocritical 
points~\cite{chen}.
The pseudo-critical excluded volume $\phi_{c}(\bar{L})$
for each pair are recorded in table~\ref{tab:phicvalues} and also appear in the graph in Fig.~\ref{fig:Fig2}, 
with the system size reciprocal $\bar{L}^{-1}$ on the abscissa.
For small system sizes, $\phi_{c}(\bar{L})$ rises, eventually saturating at $0.0317 \pm 0.0004$, the extrapolated
result for the percolation threshold.

\begin{table}
\begin{center}
\begin{tabular}{|c|c|c|c|}
\hline
$L_{1}$ & $L_{2}$  & $\bar{L}$ & $\phi_{c}(\bar{L})$   \\
\hline
5.0 & 7.5 & 12.25 & 0.03015(24) \\
6.0 & 9.0 & 7.5 & 0.03118(17) \\
7.0 & 10.5 & 8.75 & 0.03177(15) \\
8.0 & 12.0 & 10.0 & 0.03214(12) \\
9.0 & 13.5 & 11.25 & 0.03209(14) \\
10.0 & 15.0 & 12.5 & 0.03190(17) \\
12.0 & 18.0 & 15.0 & 0.03230(20) \\
13.5 & 20.25 & 16.875 & 0.03166(13) \\
14.0 & 21.0 & 17.5 & 0.03180(14) \\
16.0 & 24.0 & 20.0 & 0.03165(13) \\ 
\hline
\end{tabular}
\caption{\label{tab:phicvalues} $\tilde{\delta}$ curve intersection results with for various $\bar{L}$ values}
\end{center}
\vspace{-0.6cm}
\end{table}

\begin{figure}
\includegraphics[width=.40\textwidth]{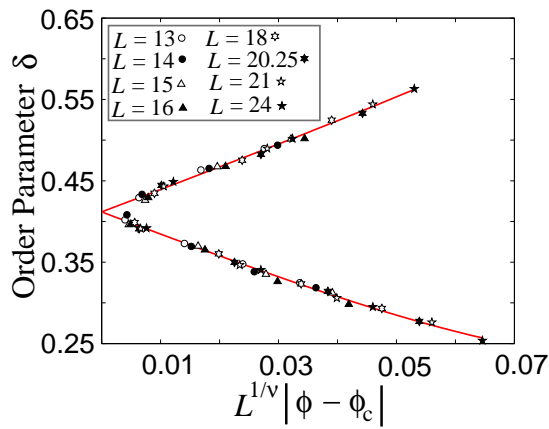}
\caption{\label{fig:Fig3} (Color online) Data collapse with $\tilde{\delta}$
data shown for various system sizes.  The solid red line is a polynomial curve
determined from a nonlinear least squares fit, and symbols represent data 
from Monte Carlo simulations.}
\end{figure}

A complementary finite size scaling technique for calculating the percolation 
threshold $\phi_{c}$, based on the data collapse phenomenon in the critical 
region, simultaneously takes into consideration results 
for a broad range of system sizes  and also yields the correlation length exponent $\nu$.
    
Data collapse graphs may be prepared by plotting on the vertical axis the quantity $\tilde{\delta}$ with 
$L^{1/\nu} | \phi - \phi_{c} |$ (with an optimal collapse for quantitatively correct values for 
$\nu$ and $\phi_{c}$) on the abscissa for various systems sizes.  The coincidence of data on 
the same curve is a manifestation of the existence of a universal singular scaling 
function for $\tilde{\delta}$ as well as an indication that $\nu$ and
$\phi_{c}$ are specified correctly.

\begin{figure}
\includegraphics[width=.49\textwidth]{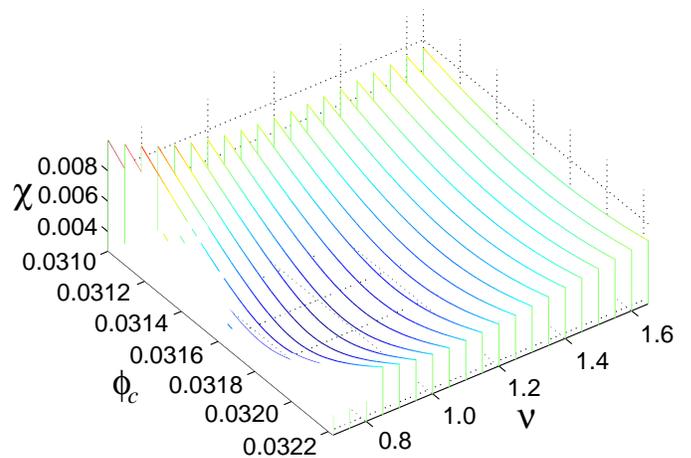}
\caption{\label{fig:Fig4} (Color online) $\chi$ plot shown with respect to the 
correlation length critical exponent $\nu$ and the excluded volume $\phi_{c}$ with 
global minimum at $\nu = 0.92$ and $\phi_{c} = 0.0317$.} 
\end{figure}

To use the data collapse as a quantitative tool, one notes that the ``v'' shaped curve in Fig.~\ref{fig:Fig3}
is essentially linear with a small degree of curvature.  To accommodate this quasi-linear variation
in a perturbative fashion,
we use a form $f(x) = A_{0} + A_{1}x +  \dots + A_{4}x^{4}$ (the results are not impacted by 
the inclusion of the quartic term, and the series is truncated at fourth order) 
with $x \equiv L^{1/\nu}(\phi - \phi_{c})$.
To assess the degree to which the points from Monte Carlo calculations lie on 
the same curve, we use 
\begin{equation}
\chi(\phi_{c},\nu,A_{0},\ldots,A_{4}) = 
\sqrt{\sum_{i}\left[ \frac{f(x_{i})-\tilde{\delta}_{i}}{\tilde{\delta}_{i}} \right]^{2}}.
\end{equation}
The quantity $\chi$ provides an overall measure of the deviation of Monte Carlo data 
from the scaling curve.
Hence, optimizing the data collapse entails minimizing $\chi$ with respect to $\phi_{c}$, 
$\nu$, and the $A_{i}$ parameters with a two tiered fit.  First, for a particular
choice of $\nu$ and $\phi_{c}$, the parameters $A_{i}$ are calculated in 
a linear least squares fit.
The appropriate values of the critical indices 
$\phi_{c}$ and $\nu$ are then determined by a stochastically driven least squares fit, 
where a sequence of attempts are made to introduce small random perturbations to $\phi_{c}$ and $\nu$, 
with the shifts accepted only if the sum of squares term $\chi$ is thereby lowered. 
Fig.~\ref{fig:Fig4} shows $\chi$ with respect to $\nu$ and $\phi_{c}$.  The global minimum 
for $\phi = 0.0317$ and $\nu = 0.92$ is, within the bounds of Monte Carlo statistical error, 
consistent with the calculation of $\phi_{c}$ based on crossings of the $\tilde{\delta}$ quantity.
Moreover, the calculated value of $\nu$ is in accord with results gleaned from high precision 
Monte Carlo calculations for discrete 3D systems (e.g. 0.875[1] for discrete percolation models
on 3D lattices with cubic symmetry~\cite{ziff}).

\section{Conclusions}

We have used a large-scale Monte Carlo calculation 
based on a dynamical exploration of void spaces to determine the 
critical properties (i.e. finding $\phi_{c} = 0.0317 \pm 0.0004$ and $\nu = 0.92 \pm 0.05$) for the percolation 
through voids surrounding randomly placed impermeable spheres.   
Two complementary finite size scaling analyses yield the same 
percolation threshold $\phi_{c}$.
Our dynamically based calculation is distinct in that it makes no approximation for the structure of 
void spaces, while also incorporating finite size effects which are exploited 
to calculate critical indices.  The result for the correlation length critical exponent $\nu$
is compatible with  the critical behavior of
discrete 3D counterparts.  

\section{Appendix: Sampling the Poissonian Distribution}

To sample realizations of disorder from the appropriate statistical distribution,
it is important to generate random configurations of spherical inclusions in an 
unbiased way; since we operate in the 
grand canonical ensemble, the number $N$ of spheres in the simulation volume
must in general vary from one sample to the next due to statistical fluctuations. 
The integer value closest to the mean occupancy $N_{\mathrm{av}} = \rho L^{3}$ is a 
convenient initial choice, and random variations in the number of spheres in the system 
are taken into consideration with a sequence of stochastically driven
attempts either to raise or lower $N$.  The latter are part of an importance 
sampling scheme similar to that used to derive the Metropolis criterion~\cite{metropolis} 
at the heart of Monte Carlo simulations that sample the Boltzmann distribution in  
the calculation of equilibrium thermodynamic variables.

To determine the probability of having $N$ spheres in the simulation volume 
$v = L^{3}$, we divide $v$ into 
$M$ subvolumes of equal size where $\Delta v = v/M$.  For large $M$ the     
likelihood of multiple occupancy in any of the subvolumes is very small
relative to the chance of having one or zero sites in a subdivision; in the     
small $\Delta v$ limit, the single occupancy probability is $\rho \Delta v$, with      
$(1 - \rho \Delta v)$ being the complementary likelihood of null occupancy.
Hence, the probability that the entire system is devoid of hopping sites is 
$(1 - \rho v/M)^{M}$, which becomes $e^{-\rho v}$ for $M \rightarrow \infty$.  
For single occupancy, adopting a prefactor $M$ to take into consideration
that the site may reside in any of the $M$ subvolumes, yields 
$M (\rho v/M)(1 - \rho v/M)^{M-1}$, which becomes $\rho v e^{-\rho v}$
in the $\Delta v \rightarrow 0$ limit.  Similar logic gives 
$P(N) = e^{-\rho v} (\rho v)^{N} /N!$ for the general case of exactly 
$N$ spheres in the simulation volume, where $N!$ compensates for 
multiple occupancy.

To generate a disorder realization, a succession of attempts
(a number of moves in the vicinity of $N_{\mathrm{av}}$ is 
sufficient to achieve ergodicity) is made to raise or 
lower the occupancy number $N$, where the choice to increase 
or decrease $N$ is randomly determined.  For increments from 
$N$ to $N+1$, the change is accepted if $X_{r} < r_{+} \equiv p(N+1)/p(N)
= \rho v/(N+1)$, where $X_{r}$ is a random number sampled uniformly 
from the interval [0,1].  Similarly, decreasing $N$ to $N-1$ occurs if 
$X_{r} < r_{-} \equiv p(N-1)/p(N) = N/\rho v$.  Finally with $N$ 
properly sampled, three Cartesian coordinates for each sphere 
are chosen independently (and at random with uniform probability density)
from the interval $[0,L]$.  

\begin{acknowledgments}
Numerical calculations have benefited from the University of Maryland 280 node High 
Performance Computing Cluster (HPCC) as well as the University of Missouri Lewis 
Cluster.   

\end{acknowledgments}


\end{document}